\newcommand{\rem}[1]{}
\documentclass[prd,twocolumn,showkeys,showpacs,preprintnumbers]{revtex4-2}
\usepackage{amsfonts,amssymb,amsmath,mathrsfs}
\usepackage{url}
\usepackage{graphicx}
\newtheorem{thrm}{Theorem}

\newtheorem{prop}[thrm]{Proposition}

\newtheorem{remark}[thrm]{Remark}
\begin{document}
%%%%%%%%%%%%%%%%%%%%%%%%%%%%%%%%%%%%%%%%%%%%%%%%%%%%%%%%%
\title[Non conservative CKG]{Non conservative conformal Killing gravity: \\
coupling the dark sector with curvature and matter}

\author{Salvatore Capozziello$^{1,2}$}
\email{capozziello@na.infn.it}
\author{Carlo Alberto Mantica$^3$}
\email{carlo.mantica@mi.infn.it} 
\author{Luca Guido Molinari$^3$} 
\email{luca.molinari@mi.infn.it}
%\affiliation{Physics Department Aldo Pontremoli,
%Universit\`a degli Studi di Milano and I.N.F.N. sezione di Milano,
%Via Celoria 16, 20133 Milano, Italy.}
\address{$^1$Dipartimento di Fisica  E. Pancini, Universit\`a degli Studi di Napoli Federico II,
Napoli\\ and Istituto Nazionale di Fisica Nucleare (INFN),  Sez di Napoli, Compl. Univ. di Monte S. Angelo,
Edificio G, Via Cinthia, 80126 Napoli, Italy.}
\address{$^2$Scuola Superiore Meridionale, Via Mezzocannone 4,
 I-80134 Napoli, Italy.}
\address{$^3$Dipartimento di Fisica  Aldo Pontremoli, Universit\`a degli Studi di Milano
and INFN,  Sezione di Milano, Via Celoria 16, I-20133 Milano, Italy}
\date{\today}

\begin{abstract} 
The so-called Harada gravity with non-conserved energy momentum
tensor is here taken into account. It includes Rastall gravity as a special case. 
The field equations are written
as Einstein equations where the source is supplemented
by a divergence-free conformal Killing tensor and a tensor proportional to the metric, 
linear in the scalar curvature and the trace of the energy-momentum tensor. 
These terms can be natural candidates for  dark sector and give rise to a coupling of
the dark sector with the matter content. The field equations are
the conformal Killing extension of Rastall gravity, and include 
unimodular gravity. In a Friedmann-Robertson-Walker (FRW)  background, the Cosmic Microwave Background (CMB)  restricts parameters so that the dark sector only couples with the trace of the stress-energy tensor.
The explicit form of the tensor for the dark sector is found, and the Friedmann and
continuity equations are presented, with a standard cosmological analysis. The sum of  energy momentum tensors of dust matter 
and of  dark fluid is conserved, and the dust energy density evolves as $\mu_M\propto a^{-3/(1+\tau)}$,
with an exponent modified by the coupling $\tau $ with the dark fluid.
\end{abstract}
%\date{25 Jan 2025}

%\subjclass[2010]{83C20 %Classes of solutions; algebraically special solutions, metrics with symmetries for problems in general relativity and gravitational theory 
%(Primary), 
%83C55, %Macroscopic interaction of the gravitational field with matter (hydrodynamics, etc.)
%83D05 %Relativistic gravitational theories other than Einstein’s, including asymmetric field theories
%(Secondary)}
%\keywords{Conformal Killing gravity; conformal Killing tensor; Robertson-Walker space-time}

\maketitle

%\keywords{\textit{Tolman-Oppenheimer-Volkoff equation, conformal Killing gravity, extended theories of gravity, conformal
%Killing tensor, static spherically symmetric spacetimes, junction conditions, Buchdahl limit}}

\section{\textbf{Introduction}}
In  \cite{Harada23} Harada proposed three theoretical criteria for a gravitational theory beyond General Relativity (GR):

1) the cosmological constant $\Lambda$ should emerge as a constant of integration;

2) the stress-energy conservation $\nabla_{j}T^{jk}=0$ is a consequence of the gravitational
field equations, rather than being assumed;

3) a conformally flat metric is not necessarily a solution in the vacuum.\\
To fulfil condition 1), he postulated the following equations
with parameters $a,b,c,d$, that generically allow for a non-vanishing
divergence of the stress-energy tensor:
\begin{align}
  H_{jkl}=&\,T_{jkl}\nonumber \\
  H_{jkl}=&a(\nabla_{j}R_{kl}+\nabla_{k}R_{lj}+\nabla_{l}R_{jk})\nonumber\\
 &+b(g_{kl}\nabla_{j}R+g_{lj}\nabla_{k}R+g_{jk}\nabla_{l}R)\label{eq:harada zero}\\
  T_{jkl}=&c(\nabla_{j}T_{kl}+\nabla_{k}T_{lj}+\nabla_{l}T_{jk})\nonumber \\
 &+d(g_{kl}\nabla_{j}T+g_{lj}\nabla_{k}T+g_{jk}\nabla_{l}T)\nonumber 
\end{align}
$R_{jk}$ is the Ricci tensor with trace $R$, $T_{kl}$ is
the stress-energy tensor with trace $T$. Note that one parameter can be arbitrarily fixed. \\
Condition 2) is satisfied if $(2a+6b)\nabla_l R = (c+6d) \nabla_l T$. As a further condition, he required that every 
solution of the Einstein equations $R_{kl}-\frac{1}{2}Rg_{kl}=T_{kl}$ solves the equations \eqref{eq:harada zero}. This implies $(a-c) \nabla_{[j}R_{kl]} + (b+c/2 +d) g_{[jk}R_{l]}=0$ (cyclic sum). 
With the arbitrary constant $a=1$, Harada chose $b=-\frac{1}{3}$, $c=1$, $d=-\frac{1}{6}$ in order to avoid an a-priori
proportionality of $\nabla R$ with $\nabla T$.  His final equations are:
\begin{align}
  H_{jkl}=&\,T_{jkl}\nonumber \\
  H_{jkl}=&\nabla_{j}R_{kl}+\nabla_{k}R_{lj}+\nabla_{l}R_{jk}\nonumber \\
 & -\tfrac{1}{3}(g_{kl}\nabla_{j}R+g_{lj}\nabla_{k}R+g_{jk}\nabla_{l}R)\label{eq:Harada first}\\
  T_{jkl}=&\nabla_{j}T_{kl}+\nabla_{k}T_{lj}+\nabla_{l}T_{jk}\nonumber \\
  &-\tfrac{1}{6}(g_{kl}\nabla_{j}T+g_{lj}\nabla_{k}T+g_{jk}\nabla_{l}T)\nonumber 
\end{align}
He then obtained the first vacuum static solution and showed, in \cite{Harada23,Harada23b}, that 
the new theory is able to face the issue of dark energy coming from the observations
of the present acceleration of the Universe \cite{Perlmutter 99,Riess 98}.

Shortly after their appearance, in \cite{Mantica 23 a-1} 
a parametrization was found, showing that the Harada equations are equivalent to
the Einstein field equations modified by a supplemental conformal Killing
tensor (CKT) that is also divergence-free: 
\begin{align}
 & R_{kl}-\tfrac{1}{2}Rg_{kl}=T_{kl}+K_{kl}\label{eq:einstein enlarged}\\
 & \nabla_{j}K_{kl}+\nabla_{k}K_{jl}+\nabla_{l}K_{jk}\label{eq:Conformal Killing mantica-1}\\
 &=\tfrac{1}{6}(g_{kl}\nabla_{j}K+g_{jl}\nabla_{k}K+g_{jk}\nabla_{l}K)\nonumber
\end{align}
where $K=g^{kl}K_{kl}$. For this reason, the theory was named conformal
Killing gravity (CKG). \\
The reformulation makes the extension of GR
explicit through the conformal Killing term, that satisfies the condition $\nabla^{k}K_{kl}=0$.\\
Readers interested in geometrical and physical applications of conformal
Killing tensors are addressed to \cite{Coll06,Kobialko22,Rani 03}.

In recent years, a number of papers have dealt with the conservative
CKG equations (\ref{eq:Harada first}).

$\bullet$ In static spherically symmetry, we mention the following ones: 
Barnes discussed vacuum solutions \cite{Barnes23a} as well as the solution 
with Maxwell source \cite{Barnes23b,Barnes 24b}; the general vacuum and electrovac solutions 
were obtained by Clem\'ent and Nouicer \cite{Clement24}.
Junior et al. 
investigated regular black-hole solutions of CKG coupled with nonlinear
electrodynamics and scalar fields \cite{Junior24,Junior 24 b,Ednaldo J 25}.
Dynamics of thin-shell wormholes derived by
two CKG black holes was pursued in \cite{Alshal 24}. 
In \cite{Mantica 24b} an anisotropic CKT was constructed,  
yielding field equations of second order: this fact considerably
simplified the derivation of solutions with respect to the Harada formulation.
In \cite{Mantica 25 } the same authors derived the analog of the
Tolman-Oppenheimer-Volkoff equation for CKG. 
In \cite{Mantica 25 b} examples of conserved
currents in static CKG were exhibited, in reply to \cite{Altas 25}. 
Kundt and wave  metrics in CKG are studied in \cite{GGurses 24,Hervik 24,Barnes24}. A discussion of
 cosmological constant is in \cite{Feng 24}. Other recent papers are
\cite{Chen 25,Clement 25,Gurses 25,Ghaffari 25,Javed 25}.

$\bullet$ In cosmology, we mention 
the conformal Killing parametrization in FRW background as a perfect fluid
$K_{jl}$ (candidate for the dark sector) \cite{Mantica 23 a-1}, with Friedmann equations
giving the same predictions for the dark fluid obtained by Harada in
\cite{Harada23b}. A $\Lambda$ term arises as integration constant,
and the field equations are second order. Vacuum cosmological solutions,
wormhole and black-hole solutions were discussed by Clem\'ent and Nouicer
\cite{Clement24}. The first attempt to obtain a quantitative derivation 
of cosmological parameters in CKG was developed in \cite{Mantica 24},
where the theoretical expression of the Hubble function versus redshift
$H(z)$ was fitted with cosmological data based on cosmic chronometers
(CC) or including baryon acoustic oscillations (CC+BAO). The dark
energy density and pressure were found to vary with the redshift,
with large dependence on the sign of the dark energy density $\Omega_D$
upon the chosen dataset.

A consistent breakthrough in the comprehension of dark energy are
the recent results of the Dark Energy Spectroscopic Instrument (DESI)
\cite{Adame 25 DESI,Adame 25 b,Abdul DESI release 2}. 
 A large literature eventually flourished, with main focus on the
persistent deviations of DESI data from standard $\Lambda$CDM cosmology
(see for example \cite{CapozzielloDESI1, CapozzielloDESI2, Malekjani 25,Barua 25,Chudaykin 24,Berti 25,Giare24a,Giare24b,Yang 25,Gialamas 25}).
Most of it investigate the evolution of dark energy by testing parametric
models for $w(a)$, like the one by Chevallier-Polarski-Linder \cite{Chevallier01,Linder 03}
or by Jassal-Bagla-Padmanabhan \cite{Jassal 05}, with experimental
data.

Walking the path of a dynamical dark energy, a detailed investigation
of cosmological parameters in CKG was pursued in the recent paper
\cite{Capozziello 25} with BAO datasets from DESI
DR 2 and the Sloan Digital Sky Survey DR16 combined with the
supernovae Ia of Pantheon+ and Union3, and CMB Planck likelihood to
fix the cosmological parameters. A non-negligible negative
value for the dark energy density $\Omega_{D}$ was obtained, notwithstanding the
matter density $\Omega_{M}$ and the present time Hubble parameter
$H_{0}$ remained consistent with the $\Lambda$CDM values. Moreover
 the deceleration parameter $q_{0}$ was derived together with the present
time dark energy equation of state $w_{0}$. Finally, we mention the paper 
\cite{Khodadi 26} where $\Omega_{D}$ was found
to be negligibly small.

Some models consider an exchange of energy momentum among
dark matter and dark energy (see for example \cite{Giare24b,Silva 25,Rugg 24,Goswami 25,Van der,Wang 24,Li 24}).
In \cite{Berti 25}, a non-parametric, model independent, reconstruction
of the dark energy density evolution, using DESI, is derived. In \cite{Odi 3-1}
a generalization of exponential $f(R)$ gravity is 
compared with $\Lambda$CDM model, by using the latest data from DESI. 

In the parametrization \eqref{eq:einstein enlarged}\eqref{eq:Conformal Killing mantica-1},
both equations $\nabla^{p}T_{pl}=0$ and $\nabla^{p}K_{pl}=0$ are
valid. To date, in all investigations regarding CKG, there is no coupling
between dark sector and matter. The purpose of this paper to investigate 
non-conservative CKG.

In Section \ref{sec:Non-conservative-harada's}, we propose a variant of \eqref{eq:harada zero}
and show that Rastall gravity is contained in it.
Then we 
write the field equations as Einstein equations where the stress-energy
tensor is supplemented by a tensor that we associate to the dark sector.
It gives rise to a coupling between the dark sector and matter and
radiation. The model is the conformal Killing extension of Rastall
gravity, and contains unimodular gravity. In Section \ref{sec:Friedmann-equations},
we present the Friedmann equations for non-conservative CKG, the continuity
equation, and the expression of the tensor representing the dark sector.
To comply with the CMB evolution, we require the dark sector to couple
only with matter and not with radiation. Furthermore, we investigate the
case with flat space curvature and neglect radiation. The (dust)
matter density evolves with the scale as $a^{-3/(1+\tau)}$, where
$\tau$ is due to the dark sector. We provide the Hubble function
$H(z)$ and the EoS of the dark fluid. In Section \ref{sec:Conclusion-and-perspectives}
we expose conclusions and perspectives. 

The Lorentzian signature is $(-+++)$.
\section{\label{sec:Non-conservative-harada's}\textbf{Non conservative Harada
equations:
the conformal Killing parametrization}}

In the initial formulation \eqref{eq:harada zero} by Harada, the equations were non-conservative, and have never been explored in their generality.\\
The choice $a = c = 1$ makes no restrictions to the
divergence of the stress-energy tensor. We then set  $b=-\frac{1}{3}(1-\sigma)$, $d=-\frac{1}{6}(1-2\tau)$, and study the non-conservative Harada equations with parameters $\sigma $ and $\tau $:
\begin{align}
H_{jkl}=&\,T_{jkl} \nonumber\\
H_{jkl}=&\nabla_{j}R_{kl}+\nabla_{k}R_{lj}+\nabla_{l}R_{jk}\nonumber\\
&-\frac{1-\sigma}{3}(g_{kl}\nabla_{j}R+g_{lj}\nabla_{k}R+g_{jk}\nabla_{l}R) \label{eq:Harada first non conservative}\\
T_{jkl}=&\nabla_{j}T_{kl}+\nabla_{k}T_{lj}+\nabla_{l}T_{jk}\nonumber\\
&-\frac{1-2\tau}{6}(g_{kl}\nabla_{j}T+g_{lj}\nabla_{k}T+g_{jk}\nabla_{l}T)\nonumber
\end{align}

Inserting the GR equations $R_{kl}-\tfrac{1}{2}Rg_{kl}=T_{kl}$ with
$-R=T$ into (\ref{eq:Harada first non conservative}), it is $(\sigma+\tau)(g_{kl}\nabla_{j}R+g_{lj}\nabla_{k}R+g_{jk}\nabla_{l}R)=0$.
Thus we have the simple statement: 
\begin{prop}
\label{prop:1} Solutions of the Einstein field equations are solutions of
the non-conservative Harada field equations (\ref{eq:Harada first non conservative})
whenever $\sigma+\tau=0$. 
\end{prop}

The contraction of \eqref{eq:Harada first non conservative} with
$g^{jk}$ gives 
\begin{equation}
\nabla^{p}T_{pl}=\nabla_{l}(\sigma R-\tau T)\label{eq:div T}
\end{equation}
The divergence of the stress-energy tensor is non-vanishing, and it  is
sourced by the gradient of its trace $T$ and of the scalar curvature
$R$. This property recalls the Rastall gravity.

The so-called Rastall gravity was introduced by Peter Rastall
in 1972 \cite{Rastall 72} as a simple generalization of the standard GR
equations with a non-conserved stress-energy tensor. The
main idea was the fact that a locally conserved stress-energy 
in flat spacetime does not necessarily imply that the same
law is valid in a curved spacetime. In this regard, the relation 
\[
\nabla^{p}T_{pl}=a_{l}
\]
was proposed, and Rastall assumed that 
$a_{l}\propto\nabla_{l}R$.
The corresponding field equation, reported in \cite{Akarsu 20},
is: 
\begin{equation}
R_{kl}-\left(\frac{1}{2}+\varepsilon\right)Rg_{kl}=T_{kl}\label{eq:rastall akarsu}
\end{equation}
with $\nabla^k T_{kl}=-\varepsilon\nabla_{l}R$.\\
In \cite{Batista 12}, it was stressed that, 
in cosmology, particle creation leads to a violation of the classical conservation
law. In this sense, Rastall's idea may be viewed as a kind of
classical formulation of that quantum phenomenon. Since the violation
of the energy-momentum conservation is connected to curvature,
the term $\varepsilon\nabla_{l}R$ was interpreted
as an effective term for particle creation.\\
Rastall gravity originated a debate about its deep nature: Matt Visser claimed in
\cite{Visser 17} that it is equivalent to Einstein gravity,
while Farhad Darabi et al. \cite{Darabi 18} argued that it is
not, with Rastall theory being {\em open to the challenges
of observational cosmology}. Anyhow, Rastall theory is largely used.
We mention for example \cite{Batista 12,Lin 20,Akarsu 20,Mohebi 25}
for cosmology, \cite{Heydardzade 16,Lin 18,Ali 26,Lobo 18,Laassiri 24,Heydardzade 17,javed 25 b}
for compact objects and black holes, \cite{Naseer 24,Naseer 24b,Naseer 25,Naseer 25 b,Naseer 25 c}
for stellar structure and \cite{Ziaie 19,Ekatria25} for gravitational
collapse.

The field equations (\ref{eq:rastall akarsu})
are not the only ones compatible with Rastall's idea $\nabla^{p}T_{pl}=a_{l}$.
As argued in \cite{Lin 18} (see eq.2.1), equations of the general form 
\begin{equation}
R_{kl}-\frac{R}{2}g_{kl}=T_{kl}- A_{kl}\label{eq:kai Lin}
\end{equation}
 where $a_{l}=\nabla^{k} A_{kl}$ are compatible and follow
the original spirit of Rastall's idea. 
Here, in non-conservative CKG, eq.(\ref{eq:div T}) gives $a_{l}=\sigma\nabla_{l}R-\tau\nabla_{l}T$.

The following result states that Rastall gravity
is contained in non-conservative Harada gravity.
\begin{prop}
A solution of the Rastall equation (\ref{eq:rastall akarsu}) solves
the non-conservative Harada equations (\ref{eq:Harada first non conservative})
if 
\begin{align}
\varepsilon=-\frac{\sigma+\tau}{1+4\tau}\label{vareps}
\end{align}
Proof:
Insert (\ref{eq:rastall akarsu}) into (\ref{eq:Harada first non conservative})
with $T=-R(1+4\varepsilon)$. The result is\\ 
$
[\varepsilon(1+4\tau)+\sigma+\tau)](g_{kl}\nabla_{j}R+g_{lj}\nabla_{k}R+g_{jk}\nabla_{l}R)=0$\\
from which $\varepsilon$ is obtained. 
\end{prop}

\begin{remark}\label{NUIS} The divergence of the non conservative
Harada equation with $T_{jk}=0$ gives the vacuum condition $\sigma\nabla_{j}R=0$.
Then, if $\sigma\neq0$, $R$ is constant. \end{remark}

A parametrization of (\ref{eq:Harada first non conservative}) involving
conformal Killing tensors similar to the one obtained in \cite{Mantica 23 a-1}
can be done here. We state the following
\begin{prop}
The non conservative Harada equations (\ref{eq:Harada first non conservative})
are equivalent to the set 
\begin{align}
 & R_{kl}-\frac{R}{2}g_{kl}=T_{kl}-\sigma Rg_{kl}+\tau Tg_{kl}+K_{kl}\nonumber \\
 & \nabla_{j}K_{kl}+\nabla_{k}K_{lj}+\nabla_{l}K_{kj} \label{eq:CKG field def}\\
 &\qquad=\frac{1}{6}\left[g_{kl}\nabla_{j}K+g_{lj}\nabla_{k}K+g_{jk}\nabla_{l}K\right] \nonumber \\
 & \nabla^{p}K_{pl}=0\nonumber 
\end{align}
Proof:
Equation (\ref{eq:Harada first non conservative}) is equivalent to
the Killing condition 
\begin{align}
 & \nabla_{j}Q_{kl}+\nabla_{k}Q_{lj}+\nabla_{l}Q_{kj}=0\label{eq:Killing}\\
 & Q_{kl}=R_{kl}-T_{kl}-g_{kl}\frac{1-\sigma}{3}R+g_{kl}\frac{1-2\tau}{6}T\label{eq:Q}
\end{align}
From (\ref{eq:Killing}) it is $2\nabla^{p}Q_{pl}+\nabla_{l}Q=0$
being $Q=g^{kl}Q_{kl}$. The new tensor $K_{kl}=Q_{kl}+\frac{1}{2}Qg_{kl}$
by definition fulfills the condition $\nabla^{p}K_{pl}=0$.\\
 Using (\ref{eq:Q}) the trace is 
\begin{equation}
K=3Q=-(1-4\sigma)R-(1+4\tau)T\label{eq:Trace Q}
\end{equation}
It is straightforward to show that $K_{kl}$ is a divergence-free
conformal Killing tensor. 
\begin{align*}
\nabla_{j}K_{kl}+\nabla_{k}K_{lj}+\nabla_{l}K_{kj} & =\frac{1}{2}\left(g_{kl}\nabla_{j}Q+g_{jl}\nabla_{k}Q+g_{jk}\nabla_{l}Q\right)\\
 & =\frac{1}{6}\left(g_{kl}\nabla_{j}K+g_{lj}\nabla_{k}K+g_{jk}\nabla_{l}K\right)
\end{align*}
With eqs (\ref{eq:Q}) and
(\ref{eq:Trace Q}) the first result in (\ref{eq:CKG field def})
is obtained. 
\end{prop}

The set (\ref{eq:CKG field def}) shows that Harada's non conservative
equations (\ref{eq:Harada first non conservative}) are equivalent
to the Einstein field equations 
\begin{align}
 & R_{kl}-\frac{R}{2}g_{kl}=T_{kl}+\Theta_{kl}\label{FIELDEQ}\\
 & \Theta_{kl}=K_{kl}-\sigma Rg_{kl}+\tau Tg_{kl}\label{eq:dark sector}
\end{align}
They are the same as (\ref{eq:kai Lin}) (i.e.
eq 2.1 in \cite{Lin 18}) with $A_{kl}=-K_{kl}+\sigma Rg_{kl}-\tau Tg_{kl}$.\\
The tensor $\Theta_{kl}$ is a natural candidate to represent the \textit{Dark Sector}
of the theory, whose form is here predicted. 

The conservation relation implied by eq.\eqref{FIELDEQ}
$$0=\nabla^{p}(T_{pl}+\Theta_{pl})=\nabla^{p}T_{pl}-\sigma\nabla_{l}R+\tau\nabla_{l}T$$
mixes dark sector and ordinary matter, where 
$K$, $R$, $T$ are related through \eqref{eq:Trace Q}.

\subsection{Conformal Killing-Rastall gravity}

If $K_{kl}=0$ it is $T=-R(1-4\sigma)/(1+4\tau)$. The field equation
\eqref{FIELDEQ} becomes Rastall's equation: 
\begin{equation}
R_{kl}-\frac{R}{2}\left[1-2\frac{\sigma+\tau}{1+4\tau}\right]g_{kl}=T_{kl}\label{eq:Rastall K=00003D00003D0}
\end{equation}
The case of non-vanishing conformal Killing tensor is thus an extension,
that we name ``conformal Killing Rastall gravity''.

\noindent We consider some special cases of this extension:

1) $\sigma=0$. The first of (\ref{eq:CKG field def}) becomes 
\begin{equation}
R_{kl}-\frac{R}{2}g_{kl}=T_{kl}+\tau Tg_{kl}+K_{kl}\label{eq:Model lambda equal 1-1}
\end{equation}
The dark sector is coupled only with the trace of the stress-energy
tensor. Eq.(\ref{eq:Model lambda equal 1-1}) does not resemble
any known theory, and will be investigated in Sect.\ref{sec:Friedmann-equations}.

2) $\sigma+\tau=0.$ The first equation of (\ref{eq:CKG field def}) becomes
\[
R_{kl}-\frac{R}{2}g_{kl}=T_{kl}+\tau(R+T)g_{kl}+K_{kl}
\]
Even with $K_{kl}=0$, it represents a broader class of the ``generalized
Rastall'' field equations proposed in \cite{Lin 20}.

 3) If $\sigma=\frac{1}{4}$ and $\tau=-\frac{1}{4}$, the field equations
assume the form 
\begin{equation}
R_{kl}-\frac{R}{4}g_{kl}=T_{kl}-\frac{T}{4}g_{kl}+K_{kl}\label{eq:traceless}
\end{equation}
It is $K=0$. Whenever $K_{kl}=0$ they represent
``unimodular gravity'' or trace-free Einstein equation. It was
 introduced in 1989 by Steven Weinberg \cite{Weinberg 89} (see
eq. 7.3) and reconsidered by G.F.R. Ellis \textit{et al.} \cite{Ellis 11}.
It has also been investigated, for example, in \cite{Corral 20,Fabris 22,Piccirilli 23,Dauda 19}.\\
When $K_{kl}\neq0$, the condition $K=0$ poses restrictions
on the existence of the conformal Killing tensor itself. For example
(see below) in a FRW background such a tensor is zero, thereby
precluding the formulation of a unimodular CKG in such background.

\section{\label{sec:Friedmann-equations}\textbf{Friedmann equations for non-conservative
CKG}}

The cosmological principle fixes the structure of the spacetime as FRW, 
\begin{align}
ds^{2}=-dt^{2}+a^2(t) \left[\frac{dr^{2}}{1-kr^{2}}+r^{2}d\Omega_{2}^{2}\right],\qquad & k=0,\pm1\label{eq:1}
\end{align}
where $a(t)$ is the scale factor. A covariant characterization is the existence of a unit time-like vector field, $u_{k}u^{k}=-1$,
such that (see \cite{Capozziello 22}): 
\begin{align}
 & \nabla_{j}u_{k}=H(g_{jk}+u_{j}u_{k}),\label{eq:torse-forming}\\
 & \nabla_{j}H=-\dot{H}u_{j}\label{EIGV}
\end{align}
where $H=\dot{a}/a$ is the Hubble parameter. Eq (\ref{EIGV}) is
equivalent to $R_{jk}u^{k}=\xi u_{j}$. The eigenvalue is related
to the cosmic acceleration: 
\begin{align}
\xi=3\frac{\ddot{a}}{a}=3(H^{2}+\dot{H}). \label{XI}
\end{align}
The Ricci tensor and the scalar curvature of the conformally flat
metric (\ref{eq:1}) are 
\begin{align}
 & R_{kl}=\frac{R-4\xi}{3}u_{l}u_{k}+\frac{R-\xi}{3}g_{kl}.\label{eq:2.3 Ricci GRW}\\
 & R=\frac{6k}{a^{2}}+6H^{2}+2\xi\label{eq:2.9 scalar R}
\end{align}
As shown in \cite{Mantica 23 a-1,Mantica 24,Capozziello 25}, in a FRW background the divergence-free conformal
Killing tensor is 
\begin{align}
K_{kl}=g_{kl}\left[\frac{5}{6}Ca^{2}-\Lambda\right]+u_{k}u_{l}\frac{Ca^{2}}{3}\label{eq:Conformal Killing GRW-1},
\end{align}
and $K=3Ca^2 - 4 \Lambda$.

\subsection{The vacuum solution}

Set $T_{kl}=0$ in the field equations \eqref{FIELDEQ}: 
\begin{equation}
R_{kl}-\frac{R}{2}g_{kl}=K_{kl}-\sigma Rg_{kl}\label{eq:vacuum}
\end{equation}
We have the following 
\begin{prop}
The vacuum solutions of non-conservative conformal Killing gravity
in FRW background are spaces of constant curvature or spaces with
constant scale factor $a$. \\

Proof:
The equation gives the trace $K=-(1-4\sigma)R$. The divergence 
gives $\nabla_{l}R=0$, that implies $\nabla_{j}K=0$. Since
$K=3Ca^{2}-4\Lambda$, it is either $\dot a=0$ ($a$ is constant), or
$C=0$. \\
For $C=0$, $K_{kl}=-\Lambda g_{kl}$ and the Ricci tensor
simplifies: $R_{kl}=\frac{R}{4}g_{kl}$. The manifold is an Einstein spacetime.
\\
 The Weyl curvature tensor in 4-dimensions is 
\begin{align*}
C_{jklm}=R_{jklm}+\tfrac{1}{2}\left(g_{jm}R_{kl}-g_{km}R_{jl}+R_{jm}g_{kl}-g_{jl}R_{km}\right)\\
-\tfrac{1}{6}R \left(g_{jm}g_{kl}-g_{km}g_{jl}\right) 
\end{align*}
Since a FRW background is conformally flat, it follows 
that the Riemann tensor is 
$$R_{jklm}=\tfrac{1}{12}R(g_{km}g_{jl}-g_{kl}g_{jm})$$
i.e. Einstein FRW spacetimes are spaces of constant curvature. According
to \cite{HE}, page 124, in a Lorentzian signature they reduce to
Minkowski, de Sitter or anti de Sitter spacetimes.  
\end{prop}

Note that Einstein spacetimes were found to be non-trivial vacuum
solutions for Rastall gravity in static spherically symmetric background
\cite{Oliveira 16}.

\subsection{The continuity equation }

Consider the stress-energy tensor of a perfect fluid $T_{kl}=(p+\mu)u_{k}u_{l}+pg_{kl}$,
with trace $T=3p-\mu$. Its divergence is 
\[
\nabla^{p}T_{pl}=3H(\mu+p)u_{l}+(\dot{\mu}+\dot{p})u_{l}+\nabla_{l}p
\]
where a dot is $u^{k}\nabla_{k}$. The conservation equation is $\nabla^{p}T_{pl}=-\nabla^{p}\Theta_{pl}=\sigma\nabla_{l}R-\tau\nabla_{l}(3p-\mu)$.
Use \eqref{eq:Trace Q} to express $R$ in terms of $T$ and $K$:
\begin{align*}
3H(\mu+p)u_{l}&+(\dot{\mu}+\dot{p})u_{l}+\nabla_{l}p\\
&=-\frac{\sigma}{1-4\sigma}\nabla_{l}K-\frac{\sigma+\tau}{1-4\sigma}\nabla_{l}(3p-\mu)
\end{align*}
Transvecting with $u^{l}$ and inserting $\dot{K}=6Ca^{2}H$ give
\begin{equation}
\dot{\mu}\left[1+\frac{\sigma+\tau}{1-4\sigma}\right]-3\dot{p}\frac{\sigma+\tau}{1-4\sigma}+3H(\mu+p)=\frac{6\sigma}{1-4\sigma}Ca^{2}H. \label{eq: continuity general}
\end{equation}
We now assume that the ordinary matter content is pressure-less matter
($p_{M}=0$) and radiation $\mu_{R}=3p_{R}$: $\mu=\mu_{M}+\mu_{R}$
and $p=\frac{1}{3}\mu_{R}$. Then 
\begin{equation}
\dot{\mu}_{M}+3H\mu_{M}+\dot{\mu}_{R}+4H\mu_{R}=-\frac{\sigma+\tau}{1-4\sigma}\dot{\mu}_{M}+\frac{6\sigma}{1-4\sigma}Ca^{2}H
\label{CONT2}
\end{equation}

In many interacting dark energy models, it is customary to assume
that both radiation and baryonic matter are uncoupled and separately
conserved. This is motivated by strong `fifth-force' observational
constraints on baryonic matter, and any significant interaction with
photons would cause deviations from photons following a geodesic path
\cite{Wang 16}.\\
In the radiation-dominated era the mass term $\mu_{M}$ was negligible.
The equation for the radiation energy density 
\[
\dot{\mu}_{R}+4H\mu_{R}= \frac{6\sigma}{1-4\sigma}Ca^{2}H
\]
has solution 
\begin{equation}
\mu_{R}=\left[\mu_{R0}-Ca_{0}^{2}\frac{\sigma}{1-4\sigma}\right]\left(\frac{a}{a_{0}}\right)^{-4}
+Ca_{0}^{2}\frac{\sigma}{1-4\sigma}\left(\frac{a}{a_{0}}\right)^{2} \label{radiation}
\end{equation}
The quadratic power is untenable on physical evidence of the past history
of the Universe, as the CMB evolution 
became suppressed in the later matter-dominated era.\\
In fact, the radiation energy density of the Universe
is due both to CMB photons  and to 
neutrinos: $\mu_{R}=\mu_{\gamma}+\mu_{\nu}$. The first
have a blackbody distribution, with energy density at any time (natural units)
\cite{Dodelson21}
\[
\mu_{\gamma}=\dfrac{\pi^{2}}{15}T_{\gamma}^{4}
\]
where $T_{\gamma}$ is the photon temperature. On the other hand
the energy density of neutrinos is computed assuming that all species are
mass-less:
\[
\mu_{\nu}=N_{eff}\dfrac{7\pi^{2}}{120}T_{\nu}^{4}
\]
 where $N_{eff}=3.046$ is the effective number of neutrino
species. The neutrino temperature
is linked to the photon temperature by $T_{\nu}=(4/11)^{1/3}T_{\gamma}$.
Then, the full radiation density is 
\[
\mu_{R}=\dfrac{\pi^{2}}{15}\left[1+\dfrac{7}{8}N_{eff}\left(\dfrac{4}{11}\right)^{4/3}\right]T_{\gamma}^{4}=\dfrac{\pi^{2}}{15}g_{\ast}T_{\gamma}^{4}
\]
where $g_{\ast}$ is the effective number of relativistic degrees of
freedom.\\
The result
\[
\dot{\mu}_{R}+4H\mu_{R}=\dfrac{4\pi^{2}}{15}g_{\ast}T_{\gamma}^{3}(\dot{T}_{\gamma}+HT_{\gamma})
\]
shows that the radiation energy density is conserved if and only if $\dot{T}_{\gamma}+HT_{\gamma}=0$,
i.e. $T_{\gamma}(t) a(t) =T_{0} a_0$,
where $T_{0}$ is the present time CMB temperature. It can be shown
that this is precisely the condition in order that the CMB 
distribution retains the blackbody form during the Universe expansion.

We then impose
$\sigma=0$ in \eqref{radiation} and in the sequel, i.e. the evolution of radiation is decoupled
from the dark sector.
\begin{align}
\mu_{R}=\dfrac{\pi^{2}}{15}g_{\ast}T_{0}^{4} \left(\dfrac{a}{a_{0}}\right)^{-4}=  \mu_{R0}\left(\dfrac{a}{a_{0}}\right)^{-4} \label{eq:radiation content independent}
\end{align}
The restriction $\sigma =0$ also avoids the nuisance that $R$ is constant in vacuo, as shown
in Remark \ref{NUIS}. 

\begin{remark}
Static spherically symmetric
spacetimes (eq. 6 in \cite{Mantica 24b})
$ds^{2}=-b^{2}(r)dt^{2}+f_{1}^{2}(r)dr^{2}+r^{2}(d\theta^{2}+\sin^{2}\theta d\phi^{2})$,
model black holes or stellar interior solutions. Two of us obtained the 
anisotropic divergence-free conformal Killing tensor
\[
K_{kl}=A(r)u_{k}u_{l}+B(r)g_{kl}+C(r)\frac{\dot u_k \dot u_l}{\dot u^r\dot u_r } 
\]
$A=\kappa_2 r^2 - 2\kappa_3 b^2(r)$, $B=\kappa_1 + 2\kappa_2 r^2 + \kappa_3 b^2(r)$
and $C=-\kappa_2 r^2$.  In the vacuum \eqref{eq:vacuum}
the same reasoning of Prop.5 leads to $\sigma \nabla_j R=0$.
Then, if $\sigma\neq 0$ it is $\nabla_j R=0$ and thus $K=(4\sigma -1)R$ is constant. 
$K= -A(r) +4B(r)+C(r) = 6(\kappa_2 r^2 +\kappa_3 b^2(r)) + 4\kappa_1$. Unless $b^2(r)\propto r^2$,
a very special case, it is  $\kappa_2 = \kappa_3=0$, so that the Ricci tensor is Einstein.\\
For $T_{kl}\neq 0$ the field equation may be rewritten as
\[
R_{kl}-\dfrac{R}{2}\left[1-\dfrac{2(\sigma+\tau)}{1+4\tau}\right]g_{kl}=T_{kl}+K_{kl}-\dfrac{\tau K}{1+4\tau}g_{kl}
\]
It is clear that a nonzero value $\sigma$ does not significantly
alter the structure of the theory. Non-conservative static
CKG will be discussed in a future paper.
\end{remark}

The equation for the evolution of the dust-matter density
simplifies: 
\begin{align}
(1+\tau)\dot{\mu}_{M}+3H\mu_{M}=0\label{eq: continuity-1-1}
\end{align}
\medskip{}
The solution with initial condition $\mu_{M}(a_{0})=\mu_{M0}$ is:
\begin{equation}
\mu_{M}(a)=\mu_{M0}\left(\frac{a}{a_{0}}\right)^{-\frac{3}{1+\tau}}\label{eq:energy density matter content-1}
\end{equation}
A non-standard exponent for $\mu_{M}(a)$ also occurs in Rastall gravity
\cite{Akarsu 20}.

 The evolution of the scalar curvature is easily obtained. 
 \begin{align}
R= (1+4\tau)\mu_{M0} \left(\frac{a}{a_0}\right)^{-\frac{3}{1+\tau}}+ 4\Lambda -3Ca_0^2 \left(\frac{a}{a_{0}}\right)^2 
\end{align}
For large values of the scale it
is driven by the dark sector.

\subsection{The Friedmann equations}

The first Friedmann equation is found by multiplying \eqref{FIELDEQ}
with $\sigma=0$ by $u^{k}u^{l}$: 
\begin{align*}
-\xi+\frac{R}{2}=\mu-\frac{1}{2}Ca^{2}+\Lambda-\tau(3p-\mu)
\end{align*}
Use \eqref{XI} and \eqref{eq:2.9 scalar R} and obtain 
\begin{equation}
3H^{2}=-\frac{6k}{2a^{2}}-3p\tau+\mu(1+\tau)-\frac{Ca^{2}}{2}+\Lambda\label{eq:First Friedmann}
\end{equation}
The second Friedmann equation is found by transvecting \eqref{FIELDEQ}
by $g^{kl}$ and is the trace condition $-R=T(1+4\tau)+K$ i.e. 
\begin{equation}
\frac{6k}{a^2}+12H^2 +6\dot{H}=-(3p-\mu)(1+4\tau)-3Ca^2 + 4\Lambda  \label{eq:Second Friedmann}
\end{equation}
\begin{remark} The first Friedmann equation \eqref{eq:First Friedmann}
and the continuity equation \eqref{eq: continuity general} (with
$\sigma=0$) imply the second Friedmann equation \eqref{eq:Second Friedmann}. 

Proof:
A dot derivative of (\ref{eq:First Friedmann}) and eq.(\ref{eq: continuity general})
give $-\frac{6k}{a^{2}}+6\dot{H}=-3(p+\mu)-Ca^{2}$. This and
the first Friedmann equation imply \eqref{eq:Second Friedmann}. 
\end{remark} 

Thus we can use the first Friedmann equation and the
continuity equation.\\
 If the energy content is made is due to dust ($p_M=0$) and radiation:
  $p=p_{R}=\frac{1}{3}\mu_{R}$ and $\mu=\mu_{R}+\mu_{M}$.  The first Friedmann equation is 
\begin{align*}
H^{2}=-\frac{k}{a^2}+\frac{\mu_R}{3}+\frac{\mu_{M}}{3}(1+\tau)
-\frac{Ca^{2}}{6}+\frac{\Lambda}{3}
\end{align*}
With the expressions \eqref{eq:energy density matter content-1} and
\eqref{eq:radiation content independent} we obtain the evolution for
the scale factor 
\begin{align*}
H^2(a) = - \frac{k}{a^2}+\frac{\mu_{0R}}{3}\left(\frac{a}{a_0} \right)^{-4}
+\frac{1+\tau}{3}\mu_{M0} \left( \frac{a}{a_0} \right)^{-\frac{3}{1+\tau}}\\
-\frac{Ca_0^2}{6} \left( \frac{a}{a_0} \right)^2 + \frac{\Lambda}{3}
\end{align*}
Divide by $H_0^2$ (the Hubble value at scale $a_0$) and obtain:
\begin{align}
\left ( \frac{H}{H_0}\right )^2 =& \Omega_R \left(\frac{a}{a_0}\right)^{-4}
+\Omega_{M}\left(\frac{a}{a_0}\right)^{-\frac{3}{1+\tau}}\nonumber \\
&+\Omega_{k}\left(\frac{a}{a_{0}}\right)^{-2}
+\Omega_{\Lambda}+\Omega_{D}\left(\frac{a}{a_{0}}\right)^{2}\label{eq:ALL H-1-1}
\end{align}
\begin{gather}
\Omega_{R}=\frac{\mu_{0R}}{3H_{0}^{2}},\;\Omega_{M}=(1+\tau)\frac{\mu_{M0}}{3H_{0}^{2}},\;\Omega_{k}=-\frac{R^{\star}}{6H_{0}^{2}a_{0}^{2}},\nonumber \\
\;\Omega_{\Lambda}=\frac{\Lambda}{3H_{0}^{2}},\;\Omega_{D}=-\frac{Ca_{0}^{2}}{6H_{0}^{2}} \label{OMEGAS}\\
\Omega_{M}+\Omega_{R}+\Omega_{k}+\Omega_{\Lambda}+\Omega_{D}=1\nonumber 
\end{gather}
While $\Omega_{M}$ and $\Omega_{R}$ are true energy densities and
are positive, on the other hand $\Omega_{k}$, $\Omega_{\Lambda}$
and $\Omega_{D}$ have geometric origin and are not necessarily positive.
\\
Since $H=\dot{a}/a$, the equation for $(H/H_{0})^{2}$ gives the
time evolution of the scale function $a(t)$. Here it is written in
terms of the redshift parameter $1+z=a_{0}/a$: 
\begin{align}
\left(\frac{H(z)}{H_{0}}\right)^{2}=\Omega_{R}(1+z)^{4}+\Omega_{M}(1+z)^{\frac{3}{1+\tau}}\nonumber \\
+\Omega_{k}(1+z)^{2}+\Omega_{\Lambda}+\frac{\Omega_{D}}{(1+z)^{2}}.\label{HSQ-1-1}
\end{align}

We may obtain the redshift $z_{eq}$ for equality of the radiation
and the (dust) matter energy densities. For $z<z_{eq}$ density perturbations are 
no longer washed out by radiation, and structure formation is possible. In $\Lambda$CDM its value
is around $3400$ \cite{Kalus26}. From \eqref{eq:radiation content independent} and \eqref{eq:energy density matter content-1}, and in view of \eqref{OMEGAS} it is 
\[
1+z_{eq}=\left(\dfrac{\mu_{M0}}{\mu_{R0}}\right)^{1-\frac{3\tau}{1+4\tau}} = \left(\dfrac{\Omega_{M}}{\Omega_{R}(1+\tau)}\right)^{1-\frac{3 \tau}{1+4\tau}}
\]
A similar relation is found in Rastall gravity (eq. 27 in \cite{Akarsu 20}):
\[ 1+z_{eq}= \left(\dfrac{\Omega_{M}}{\Omega_{R}}\right)^{1+3\varepsilon} \]
where $\varepsilon = -\tau/(1+4\tau)$ (see eq.\eqref{vareps}).

\subsection{The Dark Sector}

The right hand side of the field equation \eqref{FIELDEQ} contains
two tensors that are separately conserved: the radiation energy-momentum,
and the tensor
$T_{kl}^{M}+\Theta_{kl}=\mu_{M}u_{k}u_{l}+K_{kl}-\tau \mu_{M} \,g_{kl}$
 for dust-matter and dark sector (\ref{eq:dark sector}). 
With the expression (\ref{eq:Conformal Killing GRW-1}) of the Killing
tensor, it is 
\begin{align}
\Theta_{kl} & =u_{k}u_{l}\frac{Ca^{2}}{3}+g_{kl}\left[\frac{5}{6}Ca^{2}-\Lambda-\tau\mu_{M}\right]\label{eq:Drak Sector FRW}\\
 & \equiv(p_{D}+\mu_{D})u_{k}u_{l}+p_{D}g_{kl}\nonumber 
\end{align}
where $p_{D}$ and $\mu_{D}$ are the pressure and the energy density
of a ``dark fluid": 
\begin{align}
 & \mu_{D}+p_{D}=\frac{1}{3}Ca^{2}\label{darkfluid1}\\
 & p_{D}=\frac{5}{6}Ca_{0}^{2}\left(\frac{a}{a_{0}}\right)^{2}-\Lambda-\tau\mu_{M0}\left(\frac{a}{a_{0}}\right)^{-\frac{3}{1+\tau}}\label{darkfluid2}
\end{align}
The constant $Ca_{0}^{2}$ can be determined by the energy density
of the dark fluid $\mu_{D0}$ at scale $a_{0}$: $Ca_{0}^{2}=2(\Lambda-\mu_{D0}+\tau\mu_{M0})$.
In the equation of state $p_{D}=w_{D}(a)\mu_{D}$ is: 
\begin{align}
&w_{D}(a)=-\frac{5}{3}\\
&-\frac{2}{3}\frac{\Lambda+\tau\mu_{M0}(\frac{a}{a_0})^{-\frac{3}{1+\tau}}}{(\Lambda-\mu_{D0}+\tau\mu_{M0})(\frac{a}{a_0})^{2}-\Lambda-\tau\mu_{M0}(\frac{a}{a_0})^{-\frac{3}{1+\tau}}}\nonumber
\end{align}
with limits 
\[
w_{D}(a)\approx\begin{cases}
-5/3 & a\gg a_{0}\\
-5/3+2/3(\Lambda+\tau\mu_{M0})/\mu_{D0} & a=a_{0}\\
-1 & a\ll a_{0}
\end{cases}
\]
We explicitly stress the phantom behavior
for late times and the cosmological constant behavior for early times.

In the late Universe the contribution of radiation vanishes. For simplicity,
we also assume a flat space curvature, $k=0$. 
The equation for the Hubble parameter $H(z)$ now is: 
\begin{gather}
\left(\frac{H(z)}{H_{0}}\right)^{2}=\Omega_{M}(1+z)^{\frac{3}{1+\tau}}+\Omega_{\Lambda}+\frac{\Omega_{D}}{(1+z)^{2}}%
\label{eq:Hubble late time-1}
\end{gather}
Besides the Hubble parameter we also consider the deceleration parameter
\[
q=-\frac{a\ddot{a}}{\dot{a}^{2}}=-1-\frac{\dot{H}}{H^{2}}
\]
With $\dot{H}=\dot{z}(dH/dz)$ and $\dot{z}=-H(z+1)$, it is 
\begin{align}
q(z)=-1+\frac{1}{2}\frac{\frac{3}{1+\tau}\Omega_{M}(1+z)^{\frac{3}{1+\tau}}-2\Omega_{D}(1+z)^{-2}}{\Omega_{M}(1+z)^{\frac{3}{1+\tau}}+\Omega_{\Lambda}+\Omega_{D}(1+z)^{-2}}
\end{align}
The transition from decelerated
to accelerated phases of the Universe depends upon the
values of $\Omega_{M}$, $\Omega_{D}$ and $\tau$.\\
The present and asymptotic values  are
\[
q_0 = \frac{3}{2}\frac{\Omega_m}{1+\tau} -\Omega_D -1, \quad q_{\infty}=\frac{1}{2} - \frac{3}{2} \frac{\tau}{1+\tau}
\]
In conservative CKG: $q_{0}=\frac{3}{2}\Omega_{M}-\Omega_{D}-1$ and $q_{\infty}=\frac{1}{2}$.\\

The ``lookback time'' of a photon emitted at time $t_{e}$, at the
scale $a_{e}=a(t_{e})$, and received at present time $t_{0}$, at
the scale factor $a_{0}$, is: 
\begin{equation}
t_{L}=\int_{t_{e}}^{t_{0}}dt=\int_{a_{e}}^{a_{0}}\frac{da}{\dot{a}}=\int_{0}^{z_{e}}\frac{dz}{H(z)(1+z)}
\end{equation}
The age of the Universe (if finite) is the limit $z_c\rightarrow+\infty$
of the integral. 

A fit of the expression (\ref{eq:Hubble late time-1})
with experimental data would determine the values of $\tau$ and the densities 
$\Omega_{M}$, $\Omega_{\Lambda}$, $\Omega_{D}$. This would establish whether non conservative CKG
improves or not the conservative version, or the standard $\Lambda$CDM
model.  In  case of different behaviors it will be interesting to
evaluate the age of the Universe and the Hubble constant $H_{0}$
and other cosmological sensitive parameters.

\section{\label{sec:Conclusion-and-perspectives}\textbf{Conclusion and perspectives}}

We investigated a non-conservative version of the Harada model that extends the 
Rastall gravity and the special case of unimodular gravity. A parametrization with a conformal Killing tensor,
similar to that reported in Ref. \cite{Mantica 23 a-1}, is again possible, and reduces
the equations to second order in the metric tensor. The field equations
have the Einstein form, with the matter stress-energy tensor supplemented
by a divergence-free conformal Killing tensor and a tensor proportional
to the metric, linear in $R$ and the trace $T$ of the stress-energy
tensor. These additional tensors are interpreted to describe the dark
sector, that couples to the matter content.

In cosmology, the evolution of radiation forces to suppress the coupling of the dark sector 
with radiation, to maintain the standard evolution $\mu_{R}(a)\propto a^{-4}$.
The baryonic density is modeled as dust, and evolves
as $\mu_{M}(a)\propto a^{-3/(1+\tau)}$, where $\tau$
measures the coupling with the dark sector.
 The dark sector is described as a fluid with energy momentum $\Theta_{kl}$,
where the sum $T_{kl}^M+\Theta_{kl}$ is conserved.\\
 The evolution of the Hubble parameter $H(z)$ depends on 
 $\tau$ through the matter term. \\
The transition from a decelerated to an accelerated
phase of the Universe  depends upon the values of 
$\Omega_{M}$, $\Omega_{D}$ and  $\tau$. A statistical fit of the cosmological parameters, in particular of $\tau$,
would determine whether non-conservative CKG is more significant than
its conservative counterpart, or the standard $\Lambda$CDM model. 
It could be performed with the same analysis as in the recent paper
\cite{Chaudahary 26}, using Planck PR4 (NPIPE) CMB temperature, polarization, and lensing
observations, ACT DR6 CMB lensing, DESI DR2 baryon acoustic oscillations
measurements, and Pantheon+, DES-Dovekie, and UNion3 Type Ia supernova
compilation.

%%%%%%%%%%%%%%%%%%%%%%%%%%%%%%%%%%%%%%
\section*{Acknowledgments}
%%%%%%%%%%%%%%%%%%%%%%%%%%%%%%%%%%%%%%
S.~C. acknowledges the {\it Istituto Nazionale di Fisica Nucleare} (INFN) Sez.\ di Napoli, {\it Iniziative Specifiche} QGSKY and MoonLight-2  and the {\it Istituto Nazionale di Alta Matematica} (INdAM), gruppo GNFM, for the support. This paper is based upon work from COST Action CA21136 -- Addressing observational tensions in cosmology with systematics and fundamental physics (CosmoVerse), supported by COST (European Cooperation in Science and Technology). 
C.~A.~M. and L.~G.~M. acknowledge INFN, Sezione di Milano, {\it Iniziativa specifica} GSS. \\

\subsection*{Author contribution}  
\noindent
The three authors of this paper contributed equally to it.

\end{document}